\documentclass[runningheads]{llncs}

\usepackage[T1]{fontenc}
\usepackage{setspace}
\usepackage{graphicx}
\usepackage{amsmath}
\usepackage{amssymb}
\usepackage{xcolor}
\usepackage{hyperref}
\usepackage{caption}
\usepackage{subcaption}
\usepackage{booktabs}
\usepackage{tabularx}
\usepackage{adjustbox}

\begin{document}

\title{Draw2Learn: A Human-AI Collaborative Tool for Drawing-Based Science Learning}

\titlerunning{Draw2Learn: AI-Collaborative Drawing-Based Learning}

\author{Yuqi Hang\inst{1}\orcidID{0000-0002-4808-8481}}

\authorrunning{Y. Hang}

\institute{New York University\inst{1}
\email{yh2072@nyu.edu}}

\maketitle
\begin{abstract}
Drawing supports learning by externalizing mental models, but providing timely feedback at scale remains challenging. We present Draw2Learn, a system that explores how AI can act as a supportive teammate during drawing-based learning. The design translates learning principles into concrete interaction patterns: AI generates structured drawing quests, provides optional visual scaffolds, monitors progress, and delivers multidimensional feedback. We collected formative user feedback during system development and open-ended comments. Feedback showed positive ratings for usability, usefulness, and user experience, with themes highlighting AI scaffolding value and learner autonomy. This work contributes a design framework for teammate-oriented AI in generative learning and identifies key considerations for future research.
\end{abstract}

\section{Introduction}

Drawing is a generative learning strategy that externalizes mental models, supporting conceptual understanding through integration of verbal and visuospatial representations (Van Meter \& Firetto, 2013; Fan et al., 2023). However, drawing-to-learn effectiveness critically depends on learners' ability to create quality drawings—which requires sufficient prior knowledge, manageable cognitive load, and external guidance to support drawing construction and revision (Fiorella \& Zhang, 2018). Providing such guidance at scale remains resource-intensive, particularly for supporting metacognitive monitoring and offering timely feedback on drawing accuracy.

Recent advances in vision-language models enable AI to understand visual content, generate contextual feedback, and support creative processes in real-time (Peng et al., 2025). These capabilities create opportunities to provide scalable guidance during drawing-based learning—interpreting learner-created sketches, monitoring conceptual accuracy, and offering adaptive scaffolding. However, translating these technical capabilities into effective learning support requires careful interaction design that preserves learner agency, supports formative assessment during open-ended creation, and frames AI as collaborative teammate rather than evaluative tutor only.

We present \textbf{Draw2Learn}\footnote{Video demo: \url{https://vimeo.com/1159327790}.}, a system that explores these questions through a teammate-oriented design. The system translates learning principles into interaction patterns: generating structured drawing quests (goal-setting theory), providing optional visual scaffolds (ZPD), monitoring progress (formative assessment), and delivering multidimensional feedback (interactive tutoring feedback theory). We share formative user feedback collected during development and open-ended comments. Our contribution is a design framework grounded in learning science and preliminary user feedback indicating positive reception, providing groundwork for future research examining learning effectiveness.

\section{Theoretical Grounding}

Drawing-to-learn effectiveness critically depends on learners' ability to construct quality drawings that accurately depict structural relationships (Fiorella \& Zhang, 2018). However, students often struggle to create such drawings without appropriate guidance—they need support managing cognitive demands, leveraging prior knowledge, and monitoring drawing accuracy through comparison and revision (Van Meter \& Firetto, 2013). Providing such guidance at scale poses design challenges. We address these through a teammate-oriented design grounded in learning theories organized around three interconnected problems.

\textbf{Problem 1: Sustaining engagement during cognitively demanding tasks.} Drawing-to-learn requires sustained effort and can trigger anxiety in learners with low drawing confidence. To address motivation and affect, we integrate three frameworks. Drawing on theories of intrinsic motivation in learning (Mozelius, 2014), we structure drawing as collaborative adventure: quest-based tasks, progress visualization, and milestone celebrations transform potential anxiety into achievement-oriented engagement. Informed by \textit{goal-setting theory} (Locke \& Latham, 2002), we decompose abstract concepts into sequenced, proximal sub-goals following Bloom's Taxonomy, making complex ideas tractable through incremental success. Applying \textit{emotional design principles}, we enhance learner satisfaction through aesthetic transformations: applying artistic styles (oil painting, watercolor, anime) to finished sketches amplifies sense of accomplishment and pride in one's work, fostering positive emotional states that sustain engagement.

\textbf{Problem 2: Providing adaptive support without undermining agency.} Learners need help with drawing mechanics and conceptual gaps, but directive intervention can reduce ownership and productive struggle. Informed by \textit{Zone of Proximal Development} (Vygotsky, 1978), we frame AI as "more knowledgeable other" offering optional scaffolding—draggable helper objects available on request but not imposed. Guided by \textit{cognitive load theory} (Sweller, 1988), we design scaffolds to reduce extraneous load from drawing skill while preserving germane load for conceptual processing. Keeping scaffolds optional prevents learned helplessness and maintains learner control.

\textbf{Problem 3: Delivering formative feedback that supports rather than judges.} Feedback during generative tasks risks disrupting flow or undermining confidence if poorly timed or framed. Drawing on \textit{formative feedback design} (Shute, 2008), we structure feedback addressing four dimensions: motivational (acknowledging effort and progress), cognitive (identifying conceptual gaps), metacognitive (suggesting drawing strategies), and self-relevant (recognizing individual achievement). The language framing integrates insights from \textit{growth mindset theory} (Dweck, 2006) and \textit{self-determination theory} (Deci \& Ryan, 1985): AI avoids controlling directives ("you should draw X"), instead offering suggestions framed as collaborative support ("we could try adding..."); recognizes partial completion as legitimate progress rather than deficiency; and treats rough sketches as iterative construction rather than failed products. This approach maintains learner ownership while providing substantive guidance, instantiating the teammate metaphor through feedback that observes progress, supports next steps, and encourages before evaluating.

\section{Research Questions}

Based on the three problems identified above, we formulate research questions to assess whether users perceive the intended design solutions:\par
\textbf{RQ1 (Sustaining engagement).} Do quest-based tasks, goal scaffolding, and aesthetic transformations maintain user engagement during cognitively demanding drawing activities? Do users report positive affect and find the system attractive?\par
\textbf{RQ2 (Preserving agency).} Do optional scaffolds provide support without imposing directive control? Do users feel ownership over their drawing process and perceive the system as respecting rather than replacing their creative agency?\par
\textbf{RQ3 (Supporting through feedback).} Does multidimensional feedback encourage rather than judge, support progress without disrupting flow, and instantiate the teammate framing? Do users perceive AI as collaborative partner rather than evaluative tutor?

\section{System Design}

Draw2Learn implements the theoretical framework through quest-based collaboration (Problem 1: Engagement), learner-initiated scaffolding (Problem 2: Agency), and encouraging multidimensional feedback (Problem 3: Feedback quality). This section describes the system workflow and interface architecture.

\subsection{User Workflow}

\begin{figure}
\centering
\includegraphics[width=0.82\textwidth]{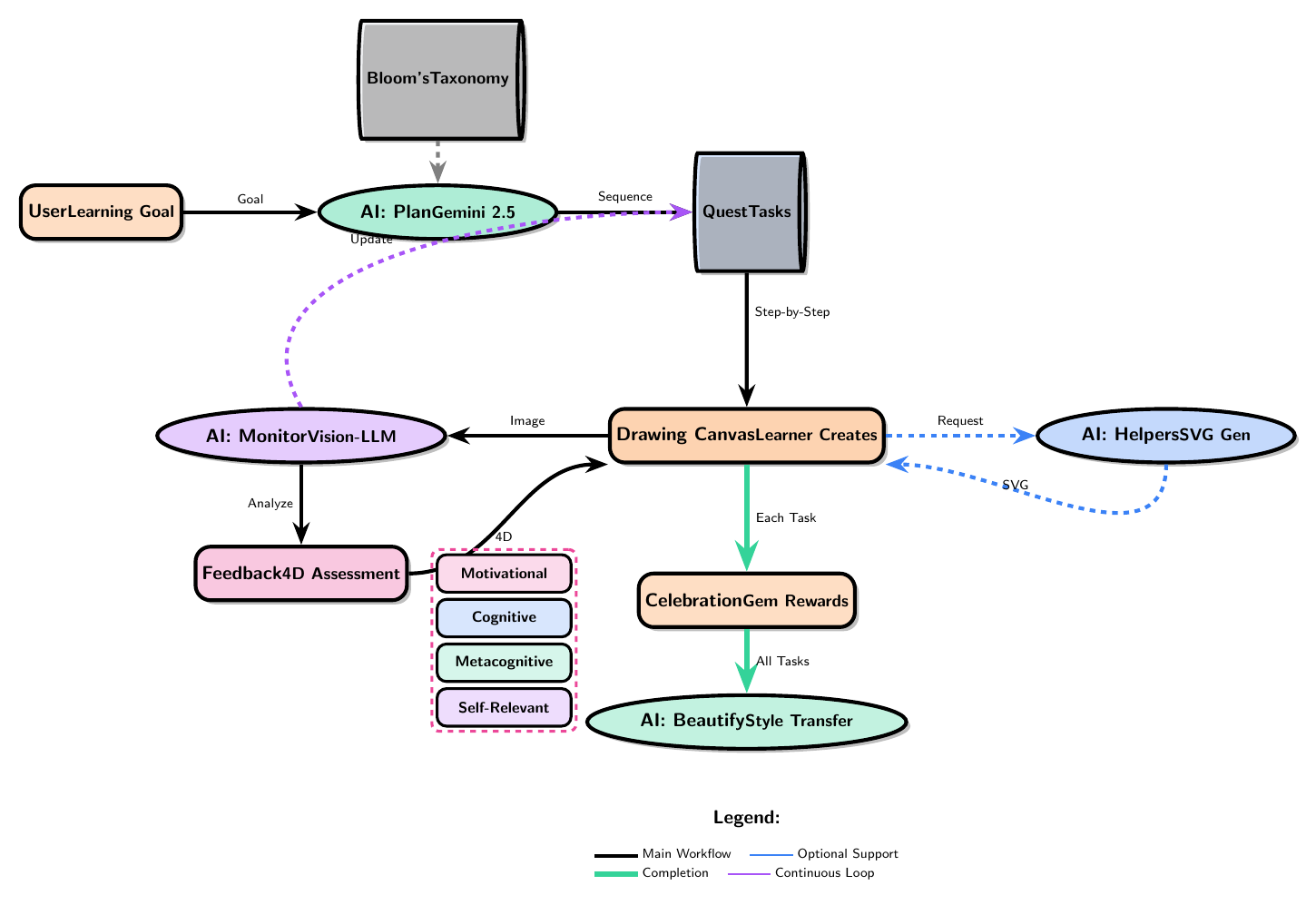}
\caption{User workflow: (1) Learner inputs learning goal; (2) AI generates quest sequence; (3) Learner draws on canvas; (4) AI monitors progress and provides 4-dimensional feedback; (5) Optional helpers available on request; (6) Gem reward upon each task completion; (7) Style transformation after completing all tasks.}
\label{fig:workflow}
\end{figure}

From the learner's perspective, Draw2Learn provides a continuous collaboration cycle (Fig.~\ref{fig:workflow}). Learners begin by entering free-text learning goals. AI transforms these into structured quest sequences following Bloom's Taxonomy progression, presenting sequential drawing tasks that scaffold from simple recall ("Draw a cell membrane") to complex synthesis ("Show how photosynthesis connects to cellular respiration"). During drawing, AI monitors canvas state through periodic vision-language analysis, generating context-aware feedback addressing motivation, cognition, metacognition, and self-relevance. Learners control task pacing and can request optional visual helpers at any point. Upon completing each task, learners receive gem rewards as immediate positive reinforcement. After completing all quest tasks, learners can apply artistic style transfer to their final work, amplifying sense of accomplishment through aesthetic transformation.

\subsection{Interface Architecture}

\begin{figure}
\centering
\includegraphics[width=0.82\textwidth]{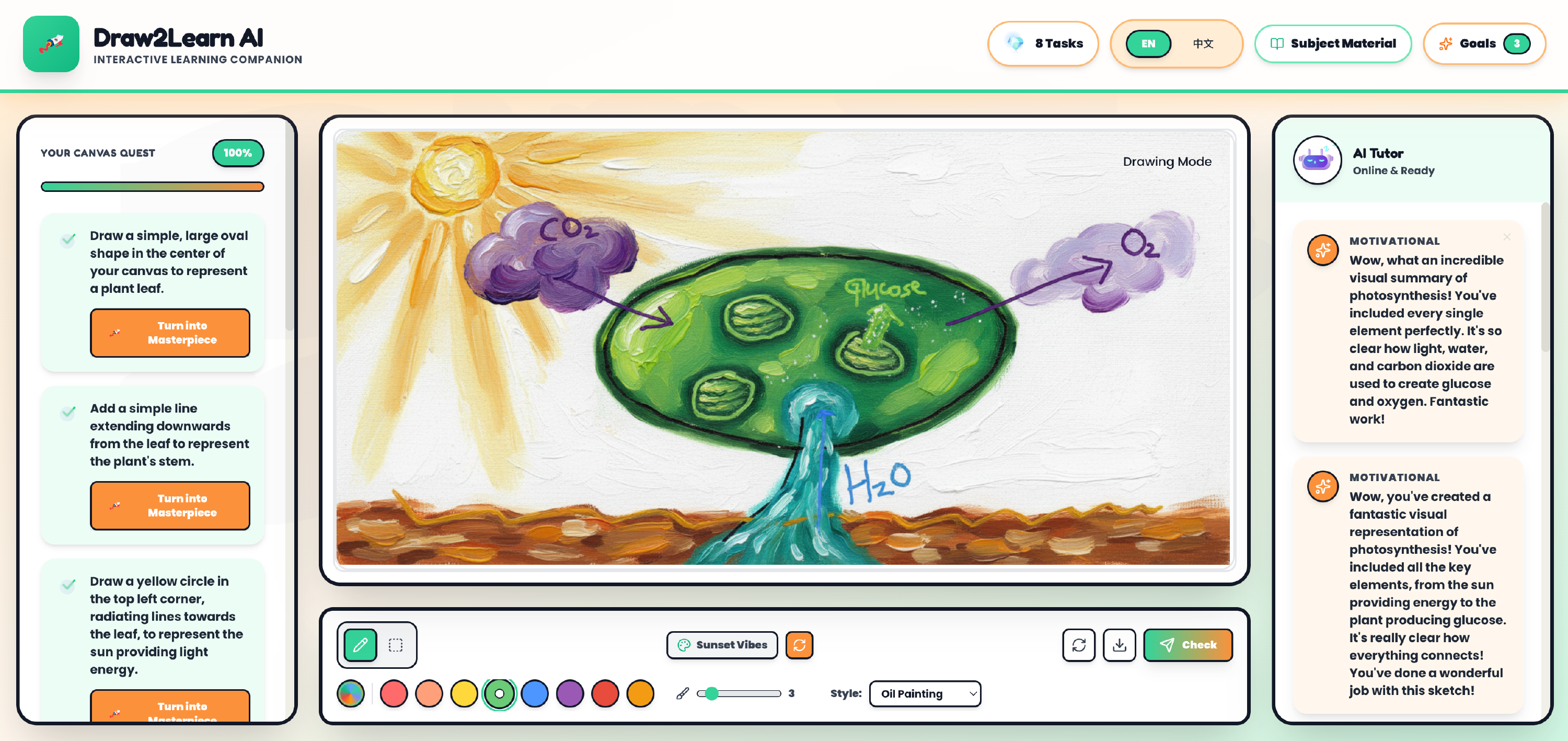}
\caption{Three-panel interface: (Left) Quest panel with progress tracking and helper buttons; (Center) Drawing canvas with style selectors; (Right) AI teammate delivering multidimensional feedback.}
\label{fig:main_interface}
\end{figure}

The three-panel interface (Fig.~\ref{fig:main_interface}) spatially distributes functionality to support flow. The \textit{left panel} ("CANVAS QUEST") presents sequential quests, progress visualization, and "helper objects" buttons for requesting visual svg objects generation that provides scaffolding opportunities without imposing them. The \textit{center panel} provides the drawing canvas with tool palette, color selectors, and style transformation options—maintaining creative control while offering aesthetic enhancement. The \textit{right panel} ("AI Tutor - Online \& Ready") delivers color-coded feedback cards addressing different dimensions, with an AI avatar signaling continuous availability—humanizing the teammate relationship while preserving learner autonomy through real-time evaluation or manual "Check" button is pressed. 

\section{Formative Feedback from Users}

\subsection{Setup}

To inform iterative design refinement, we conducted informal usability testing sessions as part of the system development process. Six individuals used the system for drawing tasks on science topics (photosynthesis, water cycle, cell structure). Following their use, they provided feedback and open-ended comments. This formative feedback process targets user experience and interaction quality (RQ1-RQ3) to guide design improvements.

\subsection{Feedback Summary}

\begin{figure}
\centering
\includegraphics[width=0.82\textwidth]{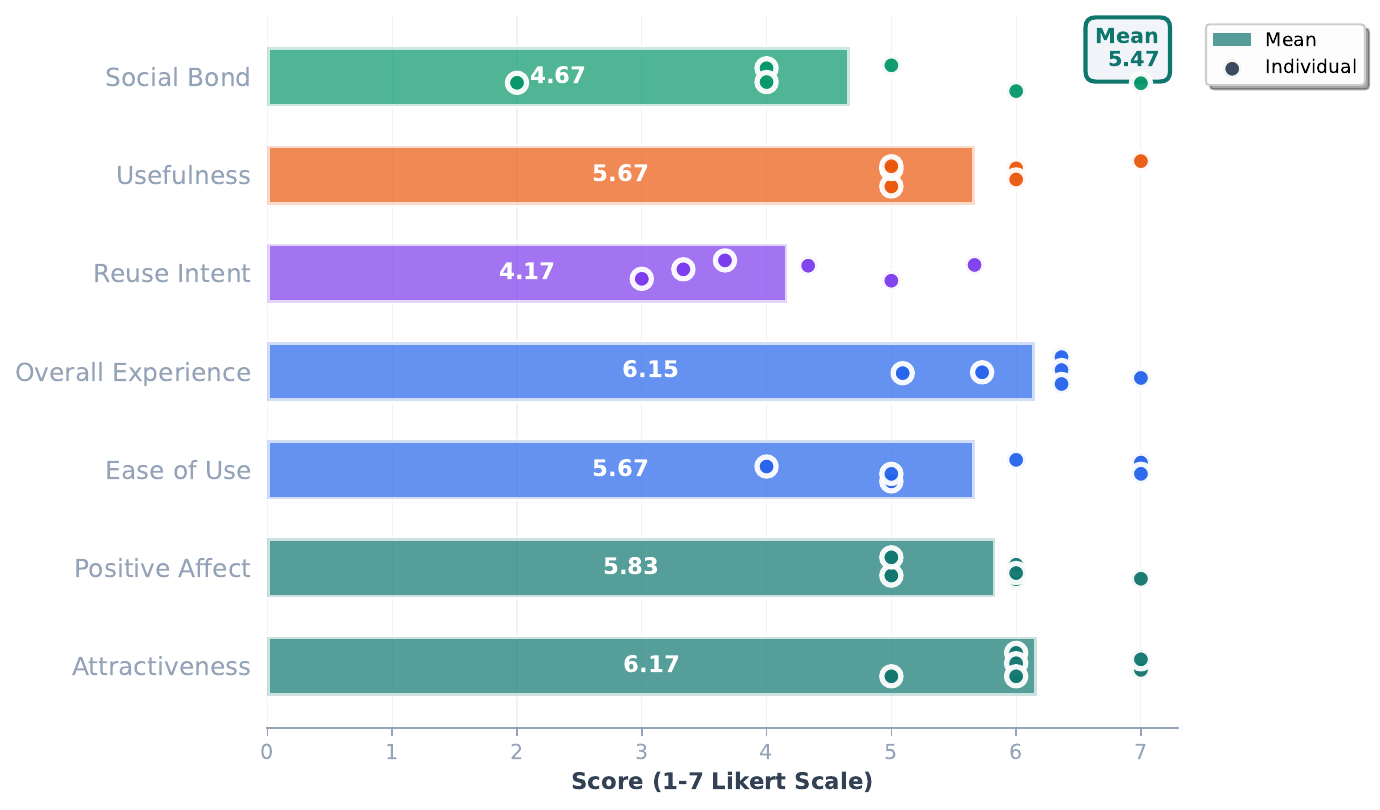}
\caption{User feedback ratings with individual data points: Positive ratings across dimensions including attractiveness (M=6.17), overall experience (M=6.15), positive affect (M=5.83), ease of use (M=5.67), and usefulness (M=5.67), with overall mean of 5.47/7.0.}
\label{fig:quantitative}
\end{figure}

User ratings (Fig.~\ref{fig:quantitative}) provide preliminary signals for RQ1-RQ3. \textit{For RQ1 (Sustaining engagement)}, high attractiveness (M=6.17/7), positive affect (M=5.83/7), and overall positive experience (M=6.15/7) suggest quest-based design and aesthetic transformations maintain motivation during cognitively demanding tasks. \textit{For RQ2 (Preserving agency)}, high usefulness (M=5.67/7) and ease of use (M=5.67/7) indicate scaffolding provides support without imposing excessive control, though direct agency measures are absent. \textit{For RQ3 (Supporting through feedback)}, moderate social bond (M=4.67/7) and reuse intent (M=4.17/7) suggest encouraging feedback is perceived and valued, though there is room to strengthen the teammate framing further. We interpret this formative feedback as design signals to guide system refinement.

\begin{figure}
\centering
\includegraphics[width=0.88\textwidth]{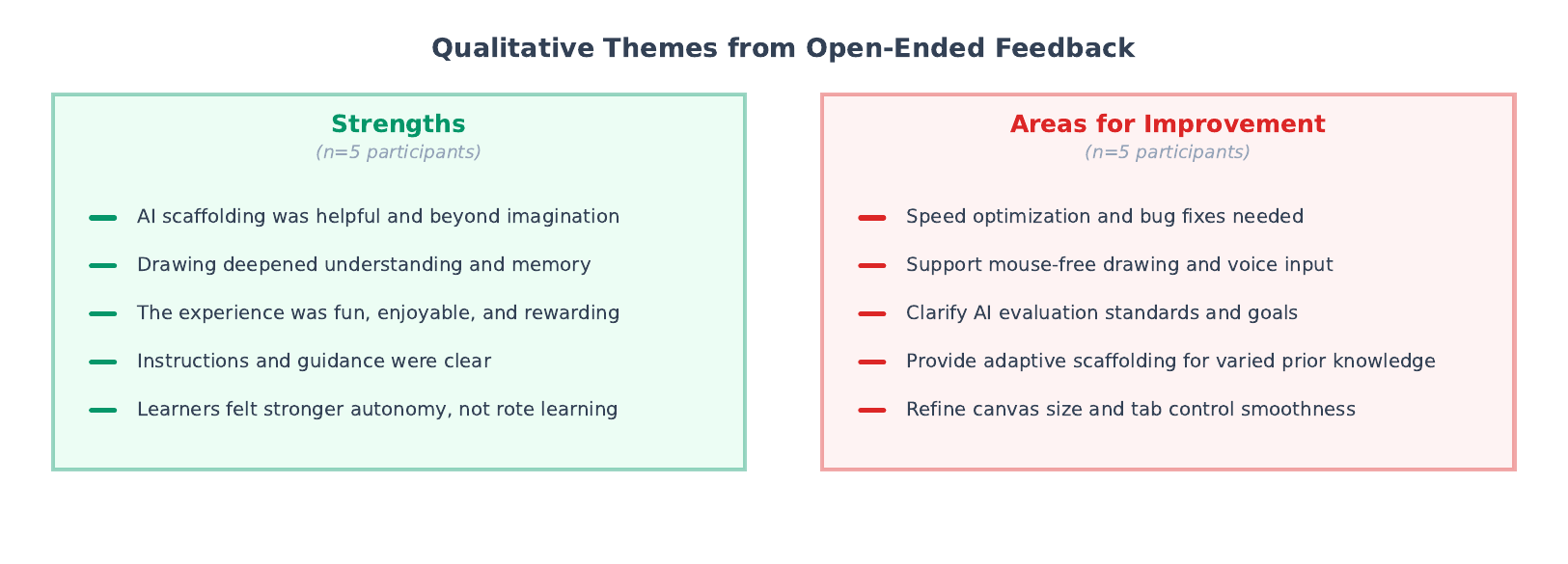}
\caption{Qualitative themes from thematic analysis: Five strengths emerged including AI scaffolding, deep learning, engagement, clear guidance, and learner autonomy. Five improvement areas include performance, input modalities, AI transparency, adaptive scaffolding, and interface refinement.}
\label{fig:qualitative}
\end{figure}

Thematic analysis (Fig.~\ref{fig:qualitative}) provides richer insights. \textit{For RQ1}, themes of "fun," "rewarding," and "deeper processing" suggest quest-based design maintains motivation. \textit{For RQ2}, "learner autonomy" and "not rote learning" indicate optional scaffolding respects creative control, though "adaptive scaffolding" requests reveal tension between support and ownership. \textit{For RQ3}, "AI scaffolding valued" and "clear guidance" suggest encouraging feedback succeeds.

\subsection{Limitations and Future Research}

This design exploration has limitations: formative feedback from six individuals provides preliminary signals rather than conclusive evidence; we did not examine learning outcomes; informal testing may not reflect authentic classroom contexts. Future work should investigate learning effectiveness through controlled experiments, classroom deployment, and examination of individual differences. 

\section{Discussion}

\textbf{Connecting theory, design, and user perception.} Formative feedback provides preliminary signals about whether design intentions reach users. \textit{For RQ1}, high attractiveness, positive affect, and themes of "fun" and "deeper processing" suggest quest-based tasks maintain motivation. \textit{For RQ2}, "learner autonomy" themes indicate optional scaffolding preserves ownership, though "adaptive scaffolding" requests reveal support-control tensions. \textit{For RQ3}, "AI scaffolding valued" suggests encouraging feedback succeeds, while moderate social bond points to opportunities for strengthening the teammate framing—users appreciate support and show initial signs of collaborative partnership that can be further developed. 

\textbf{Design contributions.} This work advances AI-supported learning in three ways. First, we introduce a \textit{teammate-oriented interaction paradigm} that reconciles AI capability with learner agency—reframing AI from directive instructor to collaborative supporter through interaction patterns that observe without intruding and scaffold without constraining. This paradigm addresses a fundamental tension in AI-supported generative learning: how to provide adaptive support without undermining the productive struggle essential for learning. Second, we demonstrate \textit{systematic operationalization of learning science principles}, bridging the gap between abstract theories and concrete interaction design. By translating motivational frameworks, cognitive theories, and feedback principles into implementable patterns (quest generation, optional scaffolds, multidimensional feedback), we provide reusable design knowledge for AI-supported generative tasks beyond drawing. Third, we present \textit{formative evidence linking design intentions to user perception}, showing that theoretically grounded interaction patterns manifest as intended user experiences—validating the teammate paradigm through convergent quantitative ratings and qualitative themes. 

\textbf{Implications.} This work demonstrates multi-framework synthesis in AI design, addressing cognitive scaffolding, affective design, and motivational support simultaneously. Key questions remain: Does the teammate paradigm transfer to other generative tasks (writing, modeling, coding)? How do individual differences moderate effectiveness? Future work should examine these questions alongside learning outcomes in authentic settings.

\section{Conclusion}

Draw2Learn explores how AI can support drawing-based learning through a teammate-oriented design that translates learning principles into concrete interaction patterns. Formative feedback during development indicates positive user experience and acceptance, with themes suggesting the design balances support with learner agency. This work contributes a design framework grounded in learning science and identifies key interaction patterns for AI in generative learning contexts. We hope this design exploration provides a useful starting point for researchers investigating AI-supported drawing-to-learn.

\end{document}